\documentclass[aps,floats,superscriptaddress,onecolumn,pre,showpacs]{revtex4-1}

\usepackage{graphicx,epsfig}
\usepackage{graphics,dcolumn,bm,epic, eepic,fleqn,float}
\usepackage{amssymb,amsmath,amsfonts,multirow,rotate,color}
\usepackage{soul}
\usepackage{wasysym}
\usepackage{dcolumn}
\usepackage{bm}
\usepackage{hyperref}

\definecolor{Myorange}{cmyk}{0,0.42,1,0}

\newcommand{\lay}[1]{^{[#1]}}

\newcommand{\beginsupplement}{%
        \setcounter{table}{0}
        \renewcommand{\thetable}{S\arabic{table}}%
        \setcounter{figure}{0}
        \renewcommand{\thefigure}{S\arabic{figure}}%
     }

\begin{document}

\title{Multiplex core-periphery organization of the human connectome}

\author {Federico Battiston}
\affiliation{School of Mathematical Sciences, Queen Mary University of London,
London E1 4NS, United Kingdom}
\affiliation{Inria Paris, Aramis project-team, 75013, Paris, France}
\affiliation{CNRS UMR-7225, Sorbonne Universites, UPMC Univ Paris 06, Inserm U-1127, Institut du cerveau et la moelle epiniere (ICM), Hopital Pitie-Salpetriere, 75013, Paris, France}

\author {Jeremy Guillon}
\affiliation{Inria Paris, Aramis project-team, 75013, Paris, France}
\affiliation{CNRS UMR-7225, Sorbonne Universites, UPMC Univ Paris 06, Inserm U-1127, Institut du cerveau et la moelle epiniere (ICM), Hopital Pitie-Salpetriere, 75013, Paris, France}

\author{Mario Chavez}
\affiliation{CNRS UMR-7225, Sorbonne Universites, UPMC Univ Paris 06, Inserm U-1127, Institut du cerveau et la moelle epiniere (ICM), Hopital Pitie-Salpetriere, 75013, Paris, France}

\author{Vito Latora}
\affiliation{School of Mathematical Sciences, Queen Mary University of London,
London E1 4NS, United Kingdom}
\affiliation{Dipartimento di Fisica ed Astronomia, Universit\`a di Catania and INFN, I-95123 Catania, Italy}

\author{Fabrizio De Vico Fallani}
\affiliation{Inria Paris, Aramis project-team, 75013, Paris, France}
\affiliation{CNRS UMR-7225, Sorbonne Universites, UPMC Univ Paris 06, Inserm U-1127, Institut du cerveau et la moelle epiniere (ICM), Hopital Pitie-Salpetriere, 75013, Paris, France}

\begin{center}
\begin{abstract}
The behavior of many complex systems is determined by a core of
densely interconnected units.  While many methods are available to
identify the core of a network when connections between nodes are all of
the same type, a principled approach to define the core when multiple
types of connectivity are allowed is still lacking. Here we introduce
a general framework to define and extract the core-periphery structure of
multi-layer networks by explicitly taking into account the
connectivity of the nodes at each layer.  We show how our method works
on synthetic networks with different size, density, and overlap
between the cores at the different layers. We then apply the method to
multiplex brain networks whose layers encode information both on the
anatomical and the functional connectivity among regions of the human
cortex. Results confirm the presence of the main known hubs, but also
suggest the existence of novel brain core regions that have been
discarded by previous analysis which focused exclusively on the
structural layer. Our work is a step forward in the identification
of the core of the human connectome, and contributes to shed light
to a fundamental question in modern neuroscience. 
\end{abstract}
\end{center}

\maketitle

\section{\textbf{Introduction}}

Network theory is a useful framework to describe many systems composed
of interacting units, from social networks to the human
brain~\cite{barabasi02,newman03,boccaletti06,latora_book}.
Real-world networks are very different from random graphs and are
characterized by the existence of typical structures from the microscopic
scale~\cite{alon02}
to mesoscopic and macroscopic scales~\cite{girvan02comm,fortunato10}.
A distinct large-scale structure is
the so-called core-periphery organisation~\cite{borgatti00}, where nodes are
partitioned into two different groups: the \emph{core}, consisting 
of a group of central and tightly connected nodes, which are usually
crucial to determine the overall behavior of the system, and the
\emph{periphery}, made by the remaining nodes.
Since the seminal paper by Borgatti and Everett~\cite{borgatti00}, the
core-periphery structure has been recognized as a fundamental property of 
complex networks~\cite{csermely13,rombach14,zhang15,barucca16}, and has been
found in several real-world systems, such as the world trade
web~\cite{fagiolo09}, many social~\cite{boyd10} and biological
networks~\cite{luo09}. A related concept is that of rich-club behavior,
where the tightly connected nodes are the network hubs, i.e. the
nodes with a large number of links
~\cite{colizza06,zhou04}. A rich-club organization has been observed
in various real-world systems, such as social, technological and biological networks
\cite{colizza06,zhou04,vaquero13, ma15}, as well as the brain
\cite{heuvel_rich-club_2011, Harriger12, VDH13, Ball14}. More recently, a 
refined version of the rich-club analysis, based not only on the
number of connections of the hubs, but also on their capability
to bridge different communities, has been shown to be relevant 
to support the integrative properties of a wide set of
networks~\cite{bartolero17}.

Rich-club and rich-core organization, associated to the efficiency in
communication and distribution of information, have been observed both in
structural and functional brain networks obtained through
image-processing from DTI or MRI data.  In the human brain it has been
conjectured that the rich cores, rather than the existence of shortest
paths, may actually be responsible for the efficient integration of
information between remote areas \cite{heuvel_rich-club_2011}, which
is a crucial prerequisite for normal functioning and cognitive
performance \cite{bullmore_complex_2009, stam_modern_2014}.  In
particular, current evidence suggests that posterior medial and
parietal cortical regions mainly constitute the core of the human
connectome, where links represent anatomical fascicles connecting
different areas.

The units of many complex systems can interact in various different
ways.  In the standard network approach, different types of
interactions are either analysed separately, losing the chance to
integrate information coming from different layers of interactions, or
aggregated all together neglecting the specific relevance and meaning
of the different types of connections. Such systems can instead be
better described as \textit{multiplex networks}, i.e. networks with
many layers, where the edges at each layer describe all the
interactions of a given type~\cite{kivela14, boccaletti14,
  battiston17challenges, dedomenico13, battiston14}.
Most of the network approaches in neuroscience have neglected the
multi-layer structure of the brain, and only recent works has focused 
on multiplex networks to merge information from different neuroimaging
modalities ~\cite{battiston17brain}, or from different
frequency components~\cite{de_domenico_mapping_2016,guillon17brain}.
At difference with other mesoscale structure, such as community
structure~\cite{mucha10comm,dedomenico15comm,battiston16comm}, the
existence and detection of core-periphery structures in multiplex
networks is a topic largely unexplored, with the exception of
approaches based on k-core decomposition~\cite{azimi14,corominas16}.

In this work we introduce a new framework to identify and
detect core-periphery organization in multiplex networks. The
method we propose works for any number of layers and is scalable
to large-scale multiplex networks, and is inspired to the algorithm 
by Ma and Mondragon for single-layer networks~\cite{ma15}.  
In the following, we first introduce the general framework and we
illustrate how the procedure works on synthetic multiplex networks with tunable
core similarity. We then apply our method to integrate
information from structural and
functional brain networks and obtain the first multiplex
characterization of the core-periphery organization of the human
brain. Our approach recovers the main hubs known in the
literature, but also allows to highlight the central role played by the regions of the sensori-motor system, 
which has been surprisingly neglected by previous studies on
core-periphery organization, despite being considered of fundamental
importance in neuroscience. 

Our research shades new light on the emergence of the core regions in the human connectome, and we hope it will spur further work towards a better understanding of the complex relationships between structure and function of the brain.

\section{\textbf{Results}}

\subsection{Extracting the rich core of a multiplex network} 

\newpage\begin{figure*}[ht!]
\begin{center}
\includegraphics[width=4in]{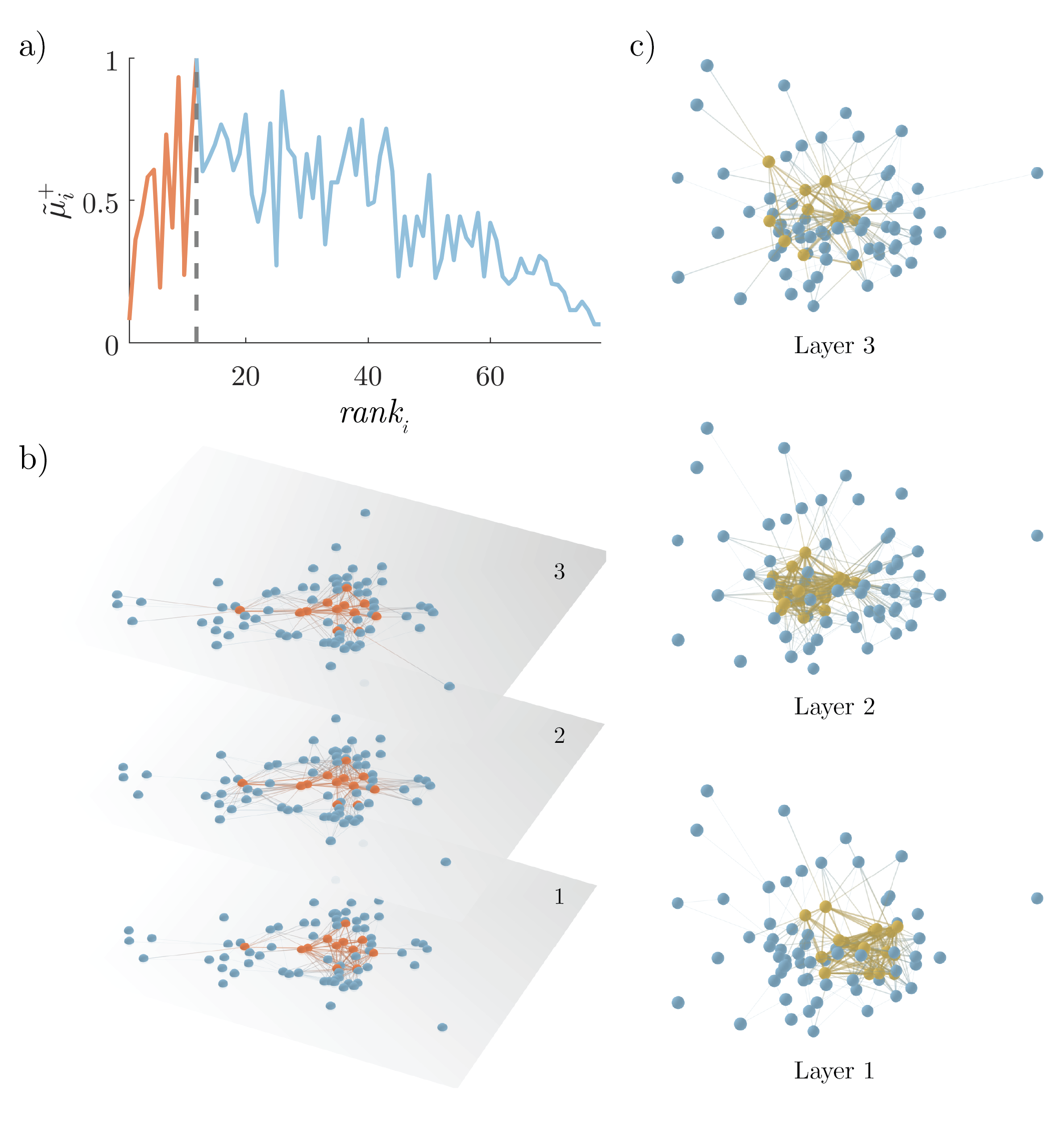}
\caption{\textbf{An illustrative example of the multiplex rich core analysis.}  
In panel (a) we show the curve $ \tilde \mu_i^+=\mu_i^+/\rm{max}(\mu_i^+)$ as a function of $rank_i$ for the Top Noordin Terrorist network, a 
multiplex social network with $N=78$ individuals, $M=3$ layers and $K\lay 1=259$, $K\lay 2=437$ and $K\lay 3=200$, where $K\lay \alpha$ indicates the number of links on layer $\alpha$. All nodes from rank equal to 1 up to the node with maximum $\tilde \mu^+$ are part of the core of the multiplex network, which is shown in red color in panel (b). The cores obtained at each layer by a standard single-layer rich core 
analysis are reported for comparison in panel as yellow nodes (c). 
}
\label{fig:fig_terr} 
\end{center}
\end{figure*}

Let us consider a multiplex network described by a vector of
adjacency matrices $\mathcal M=\{A\lay 
1, \ldots, A\lay M\}$, where all interactions of type $\alpha$,  
$\alpha=1,\ldots,M$, are
encoded in a different layer described by the adjacency matrix
$A\lay \alpha$, 
In our method to detect the core-periphery structure of a multiplex network, 
we first compute the multiplex degree vector $\bm k_i=\{k_i\lay 1, \ldots, k_i\lay
M\}$ of each node $i$ ~\cite{battiston14}. From now on, we refer to $k_i^{[\alpha]}, \alpha
=1, \ldots, M,$ as the \textit{richness} of node $i$  at layer $\alpha$. 
Notice
that this is the simplest way to define the richness of a node, and
different measures of richness, such as other measures of node
centrality, can be as well used.  For each layer $\alpha$, we then divide
the links of node $i$ in those towards lower richness nodes, and those towards
higher richness nodes, so that we can decompose the degree of
node $i$ at layer $\alpha$ as $k_i^{[\alpha]} = k_i^{[\alpha]-} +
k_i^{[\alpha]+}$.
Finally, the {\em multiplex richness} $\mu_i$ of node $i$ is obtained
by aggregating single-layer information: 
\begin{equation}
\mu_i = \sum_{\alpha=1}^M c\lay \alpha k_i\lay \alpha.
\label{eq:mu}
\end{equation}
where  the coefficients $c\lay{\alpha}$ modulate the relative relevance of each layer and can, 
for instance, be determined by exogenous information. In analogy to the single-layer case, we define 
the multiplex richness of a node towards richer nodes as:
\begin{equation}
\mu_i^+ = \sum_{\alpha=1}^M c\lay \alpha k_i^{[\alpha]+}.
\label{eq:muplus}
\end{equation}
In the most simple set-up we can assume $c\lay \alpha=c=1/M$ $\forall \alpha$.  More
general functional forms to aggregate the contributions from different layers, giving
rise to alternative measures of $\mu_i$ and $\mu_i^+$, are discussed in
the Methods section.

The nodes of the multiplex are ranked according to their richness
$\mu$, so that the node $i$ with the best rank, i.e. $rank_i=1$, is the
node with the largest value of $\mu$, the node ranked 2 is the one with the second largest value of $\mu$, and so
on. We then plot for each node $i$ the value of $\mu_i^+$ as a function of
$rank_i$. Finally, the maximum of $\mu_i^+$ is evaluated as a function
of the rank. All nodes with rank to the left of such a value are part of
the multiplex core, whereas the remaining ones are part of the
periphery. 
As an illustrative example of how our method works in Fig.~\ref{fig:fig_terr} we report the curve 
$\mu_i^+$ as a function of $rank_i$ obtained in the case of the Top Noordin Terrorist network, a multiplex network of $N=78$ individuals with three layers (encoding information about mutual trust, common operations and exchanged communication between terrorists), which has been used as a benchmark to test measures and models of multiplex networks~\cite{battiston14}. 
Coefficients $c\lay{\alpha}$ were chosen, in this case, to be inversely proportional to $K\lay \alpha$ to compensate for the different 
densities of the three layers. The resulting multiplex rich core integrates information from all the layers and 
looks different from the rich cores obtained at each of the three 
layers by a standard single-layer rich core analysis. 

\subsection{Testing the method on multiplex networks with tunable core similarity}

A network with a well defined 
core-periphery structure has a high density of links among core nodes. With a suitable labeling of the nodes, the adjacency matrix 
of the network can be decomposed into four different blocks: a dense diagonal block  
encoding information on core-core links, a sparser diagonal block describing links among peripheral nodes, 
and two off-diagonal blocks encoding core-periphery edges. 
The key feature of such block-structure is that $\rho_1 \gg \rho_3$, i.e.  
the density $\rho_1$  of the core-core block is much higher than that of the 
periphery-periphery block, $\rho_3$. As first noted by Borgatti and Everett~\cite{borgatti00}, the density $\rho_2$  
of the off-diagonal blocks is typically not a crucial factor to characterise a core-periphery structure.

In order to test how our method works on multiplex networks with
different structures, we have introduced a model to produce synthetic
multiplex networks with tunable core similarity. In particular, we
have constructed networks in which, each of the $M=2$ layers has
$N=250$ nodes and $N_c=50$ of them belong to the core. Each layer has
the same average node degree $\langle k \rangle = 10$, and the same
set of parameters $\rho_1 > \rho_2 > \rho_3$ to describe its
core-periphery structure. Our model allows to control the number of
nodes that are both in the core of layer 1 and in the core of layer 2.
(see Methods for the choice of the parameters and related discussion).

In order to quantify the similarity among cores at different layers,
we introduce the core similarity $S_c\lay{\alpha}$ of layer $\alpha$
with respect to the other layers as:
\begin{equation}
 S_c\lay{\alpha}=\frac{1}{(M-1)}\sum_{\beta \neq \alpha}^{M}  
\frac{I_c\lay{\alpha \beta}}{N_c\lay{\alpha}},
\label{eq:S_c} 
\end{equation}
where $I_c\lay{\alpha \beta}$ is the number of nodes which belong to the
core of both layer $\alpha$ and layer $\beta$, whereas $N_c\lay{\alpha}$ is the size of the core at layer
$\alpha$.  The core similarity
$S_c\lay{\alpha}$ ranges in $[0,1]$. When layer $\alpha$ does not
share core nodes with any other layers we have $S_c\lay{\alpha}=0$,
when all its core nodes also belong to the cores of the other layers
$S_c\lay{\alpha}=1$, and when on average only half of them are part of
the cores on each other level $S_c\lay{\alpha}=1/2$. The average core
similarity of the multiplex can then be computed as
$S_c=(1/M)\sum_{\alpha=1}^M S_c\lay{\alpha}$.

\begin{figure*}[ht!]
\begin{center}
\includegraphics[width=5in]{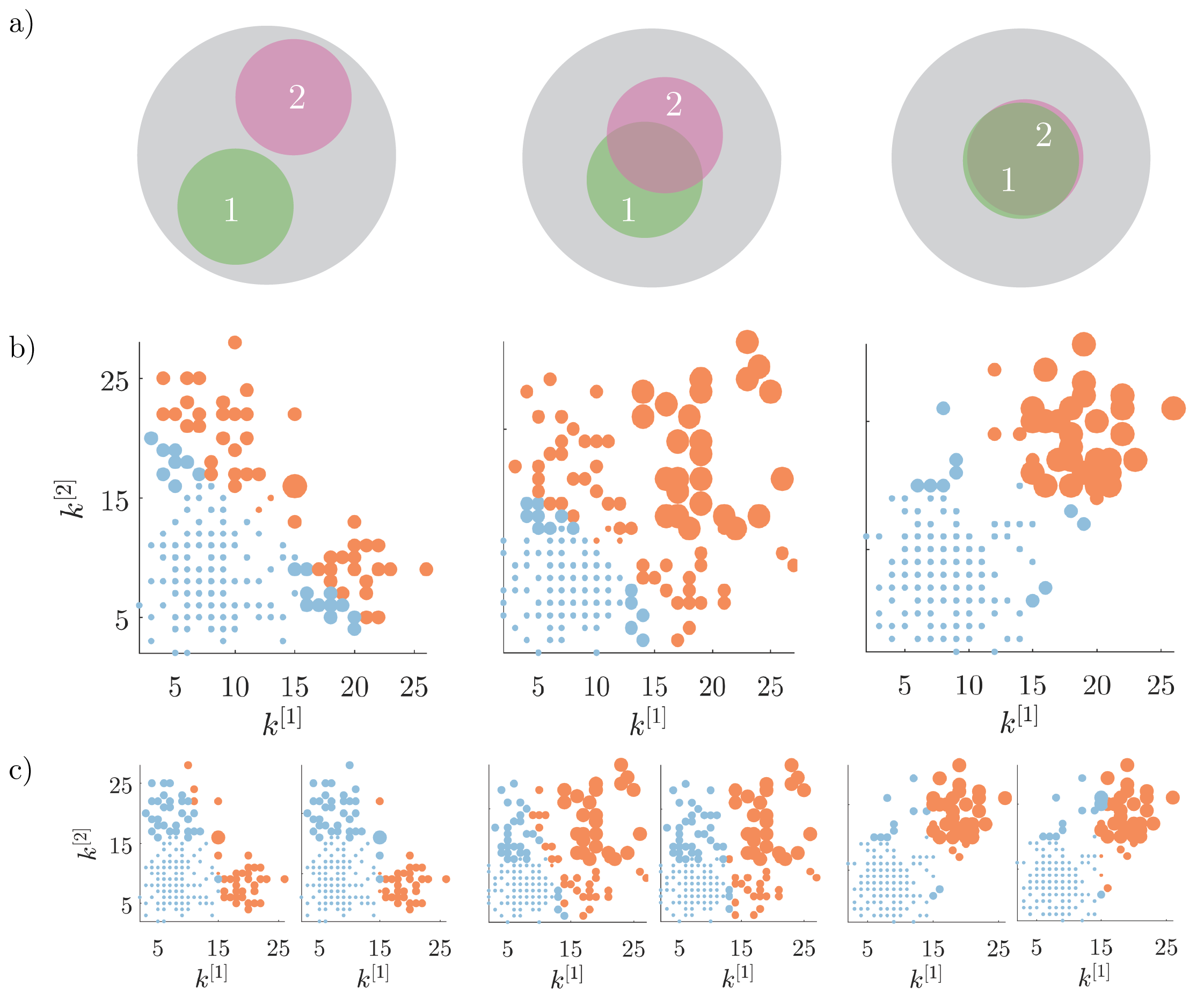}
\caption{\textbf{Core-periphery structure in synthetic multiplex networks with different core similarity.} In panel (a) we sketch multiplex networks with $M=2$ layers, $N=250$ nodes and different levels of core similarity, namely $S_c=0$ (left column), $S_c=1/2$ (central column) and $S_c=1$ (right column).  In panel (b) the nodes are placed in a two dimensional plane according to their degree at each layer. 
The size of each dot is proportional to the multiplex richness $\mu_i$  of the node, (with $c\lay{1}=1,c\lay{2}=0$). Nodes belonging 
to the multiplex cores are usually placed in the right-top corner of the plots and are colored orange, while 
the multiplex periphery is in blue. In panel (c) we report results obtained for two cases with $c\lay{1}\neq c\lay{2}$,
namely:  $(c\lay{1}=0.75,c\lay{2}=0.25)$  where the core is biased towards the important nodes of the first layer 
(left), and   $(c\lay{1}=1,c\lay{2}=0)$, where the core corresponds to the core of the first layer (right).}
\label{fig:fig_sim}
\end{center}
\end{figure*}

In Fig.~\ref{fig:fig_sim} we show results for three multiplex networks
with different core similarity. In Fig.~\ref{fig:fig_sim}(a) we
consider a multiplex with $S_c=0$. The cores of the two layers are not
overlapping, with many nodes with high degree in one layer having low
degree in the other one. In this case, the multiplex core of the
system is formed by those nodes with sufficiently high multiplex
richness. In Fig.~\ref{fig:fig_sim}(b) we consider a multiplex with
$S_c=1/2$. Half of the core nodes are shared with the other level of
the system, and half are typical of each level. The block
representation of the two layers is partially overlapping, and the
nodes are spread uniformly over the $(k_i\lay 2$ vs $k_i\lay 1$ plane.
The multiplex core of the system in this case is formed by nodes which
are part of the core on both layers, but also by nodes scoring
extremely high in one layer, despite being periphery in the other one.
At last, in Fig.~\ref{fig:fig_sim}(c) we consider a multiplex with
$S_c=1$. The block structure of the two layers is equivalent, the node
degrees $k\lay 1$ and $k\lay 2$ at the different levels are correlated
and most of the nodes belonging to each core are in the multiplex core
of the system.

\newpage

\subsection{Merging structure and function to extract the connectome's core} 

We have applied our method to investigate the human connectome by
considering, at the same time, structural and functional
information. We have therefore constructed a multiplex human
connectome network formed by one structural layer and one functional
layer.  The two layers were obtained by first averaging brain
connectivity matrices estimated respectively from diffusion tensor
imaging (DTI) and functional magnetic resonance imaging (fMRI) data in
$171$ healthy individuals. Each of the two layers is then thresholded
by fixing the average node degree $\langle k \rangle$ (Methods).

\begin{figure*}[ht!]
\begin{center}
\includegraphics[width=7in]{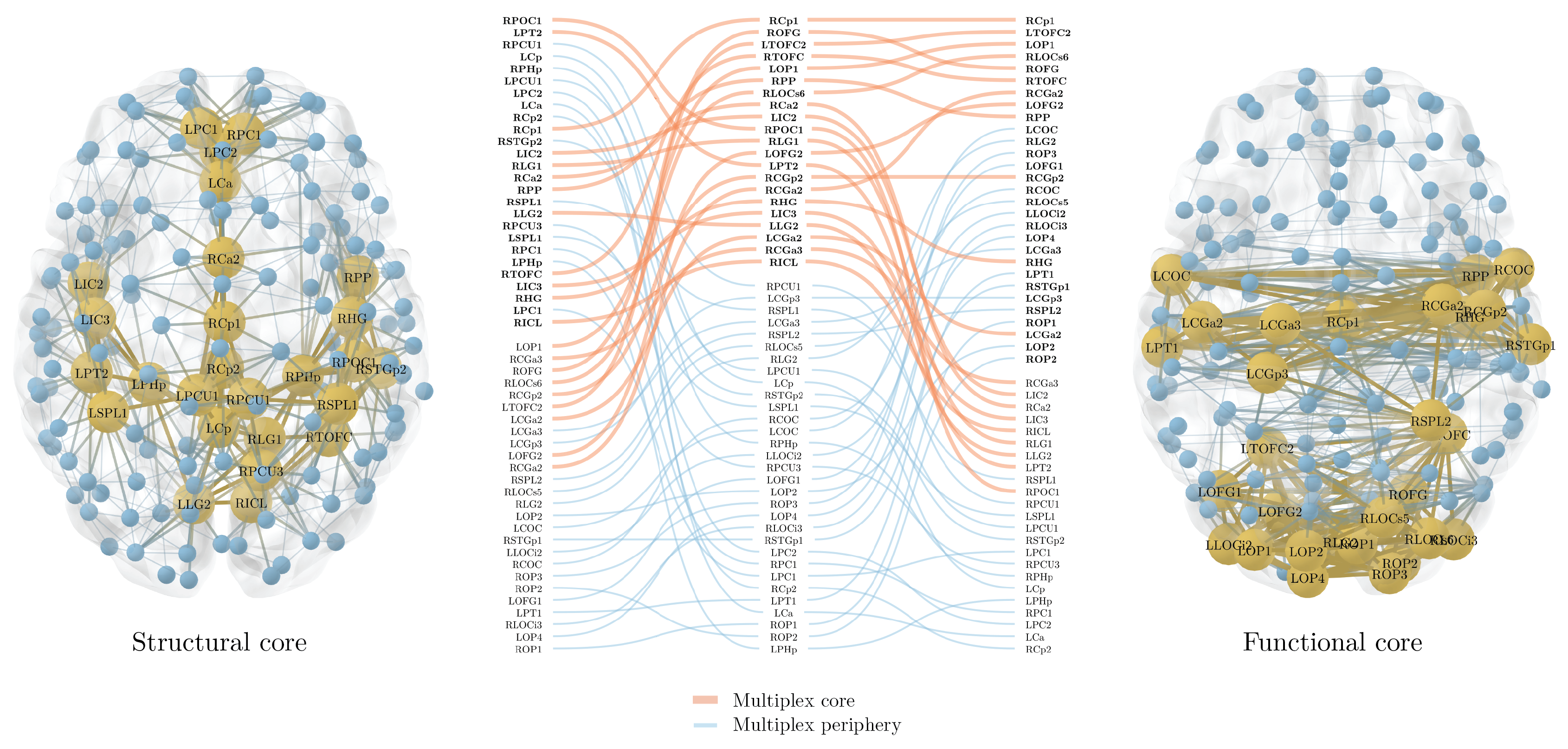}
\caption{\textbf{Extracting the multiplex core of the human brain from
    structural and functional information.} The structural and
  functional brain networks filtered with an average node degree
  $\langle k \rangle=7$ are shown respectively on the left and right
  side. They are represented from above with the frontal lobe pointing
  upward. The position of the nodes corresponds to the actual location
  of the brain regions of interests (ROIs, Table S1). 
  Yellow and large nodes represent the brain regions belonging to the core according to the
  standard single-layer method. Blue and small nodes code for the ROIs
  in the periphery. Links are yellow and thick if they connect two
  ROIs in the core, while they are blue and thin if they connect two
  peripheral nodes.  In the central part of the figure, the ROIs are
  ranked from top to bottom according to their richness in the
  structural (left column), functional (right column) and multiplex
  network (central column). In each column, the labels in bold/normal
  font stand for the ROIs that are in the core/periphery. For the sake
  of simplicity, only ROIs that are at least in one core (structural,
  functional or multiplex) are listed in the three columns. Red/blue
  and thick/thin lines identify ROIs that go into the
  core/periphery according to the multiplex approach. }
\label{fig:fig_k7}
\end{center}
\end{figure*}

In Fig.~\ref{fig:fig_k7} we report the two cores found by
separately analysing the two layers, together with the 
multiplex core obtained by our method. The figure refers to the case of 
a representative threshold corresponding to an average node
degree $\langle k \rangle=7$. 
We notice that the cores of the structural and functional layers are
only partially overlapping, with a value of cover similarity 
of $S_c=0.15$. Interestingly, ventral brain areas mainly
constitute the structural core, while more dorsal regions appear in
the functional core.

Information at the two layers is integrated by our method to extract
the multiplex core of the human connectome. 
Brain regions of interest (ROIs, Table S1) that
are in the core of both structural and functional layers also tend to be in the core of the multiplex. Instead, ROIs being in the periphery
of both layers tend to be excluded from the multiplex core. However, exceptions may exist depending on the multiplex richness of the nodes. For example, the posterior part of the right precentral gyrus (RCGa3), which is in the periphery of both the structural and functional layer, is eventually assigned to the multiplex core, because of its relatively high rank score in the two layers.  
The situation appears even less predictable for ROIs that are in the core of one layer and in the periphery of the other layer. Only occasionally these will belong to the multiplex core. 
This is the case, for example, of the anterior part of right precentral gyrus (RCGa2) which exhibits a relatively low structural richness but high functional richness, i.e. ranked seventh in the functional core.

\subsection{Revealing new core regions of the human brain}

We have extracted the multiplex core-periphery structure of the human brain for the
full range of available thresholds $ \langle k \rangle =1, 2, \ldots, 120$ (Methods).
In this way, we have been able to calculate the \emph{coreness} $C_i$ of each node $i$, 
defined as the normalised number of thresholds at which the corresponding ROI is present in
the rich core. This allows us to
rank ROIs according to their likelihood to be part of the multiplex core and to compare these 
to the rankings obtained separately for the structural and functional layers.  
We note that the same approach of investigating the persistence across a set of different 
filtering thresholds can be applied to any node property. This can turn useful for statistical 
validation in the case no threshold is universally accepted, as often happens for brain networks.

Parietal (pre/cuneus PCU/LOC, superior parietal lobe SPL), cingulate
(anterior Ca, posterior Cp), temporal (superior temporal gyrus),
insular (insular cortex IC), as well as frontal ROIs (paracingulate
PC) mainly constitute the structural core, as shown in Fig.~\ref{fig:fig_kall_struct}.  
While some overlap exists between the structural and the functional cores, the latter rather tends 
instead to include
occipital (occipital fusiform gyrus OFG, temporo-occipital fusiform
cortex TOFC) and central (pre/post central gyrus CGa/CGp) ROIs and,
notably, to exclude regions in the frontal lobe (top $25\%$ ROIs,
Fig.~\ref{fig:fig_kall_func}).  

\begin{figure*}[ht!]
\begin{center}
\includegraphics[width=7in]{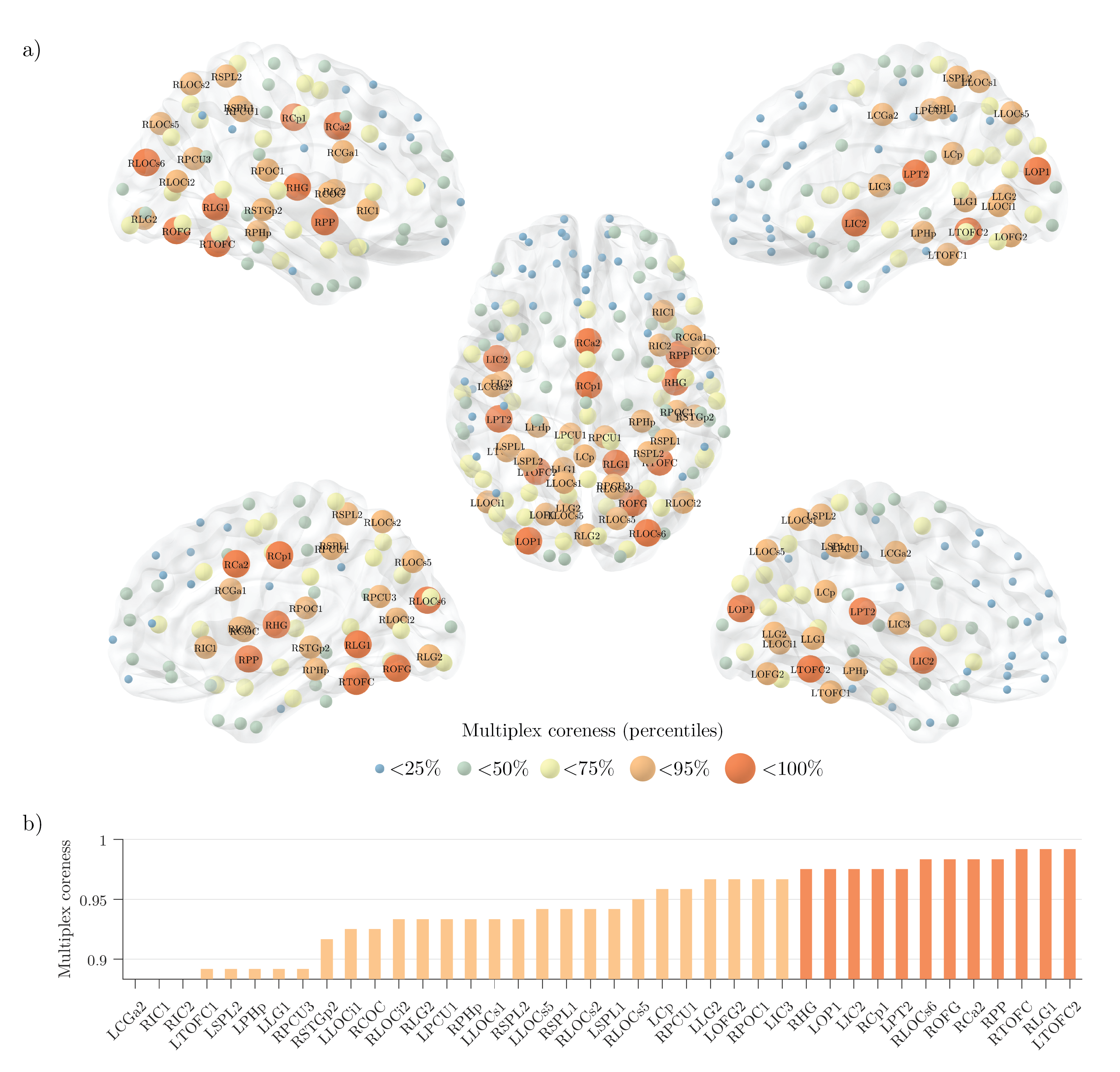}
\caption{\textbf{The multiplex core of the human connectome.} 
Panel (a) shows the human brain, where regions of interest (ROIs) are highlighted based on their multiplex coreness. The color and size of the nodes are associated to the percentiles of multiplex coreness in each brain region, so that core nodes are larger in size and coloured in red. Left side shows the lateral view of the left hemisphere (top=dorsal, bottom=ventral). Right side shows the lateral view of the right hemisphere (top=dorsal, bottom=ventral). In the middle, the brain is shown from above, with the frontal lobe pointing upward. 
In panel (b) we report the ROIs corresponding to the $25\%$ highest values of multiplex coreness. The color follows the same legend as in panel (a). }
\label{fig:fig_kall}
\end{center}
\end{figure*}

Fig.~\ref{fig:fig_kall} shows the coreness of the multiplex network.
As expected, ROIs that are peripheral (i.e., low coreness) in both
layers are also peripheral in the multiplex, while ROIs with both a
high structural and high functional coreness are typically observed in
the multiplex core (e.g., TOFC, OFG, Ca, Cp).  Interesting behaviors
emerge for those regions typically characterized by high coreness in
one layer and low coreness in the other layer.  In fact, some of these
ROIs are part of the multiplex core, while others are usually found in
the multiplex periphery, as shown Fig. \ref{fig:fig_scatter}(a).  For
regions with different assignment at the two layers, we note that the
main contribution to the multiplex richness $\mu_i$ comes from the
richness at the layer where node $i$ was identified as core. However,
not only the average richness at the layer where the node is core is
higher than the one at the peripheral layer, but so are fluctuations
around the mean. As a consequence, among regions in the structural
core but in the functional periphery, those with relatively higher
structural richness (degree), such as precuneus PCU, insular cortex IC
and posterior cingulate Cp, tend to join the multiplex core no matter
the exact value of their functional richness (upper right corner of
Fig. \ref{fig:fig_scatter}(a)). Conversely, ROIs with relatively lower
structural degree are usually peripheral in the multiplex, and
typically located in the pre-frontal cortex PC and frontal lobe FP
(lower right corner of Fig. \ref{fig:fig_scatter}(a)), as supported by
Fig.~\ref{fig:fig_scatter}(b,c). Similarly, for regions in the
functional core, those with relatively higher functional degree, such
as precentral gyrus CGa and central operculum COC, tend to join the
multiplex core (upper left corner of
Fig. \ref{fig:fig_scatter}(a)). In contrast, ROIs with relatively
lower functional degree, are mostly peripheral in the multiplex, and
are located in the parietal operculum POC and superior frontal gyrus
SFG (lower left corner of Fig. \ref{fig:fig_scatter}(a)).

\section{\textbf{Discussion}} 

Most networks have a core, a subset of central, tightly interconnected
nodes at the heart of the system, which often are the primary
responsible for the emergence of collective behaviors. Finding the
core-periphery structure of a network is a problem of crucial
importance in network science, and a variety of different algorithms
have been proposed to define and extract the core in single-layer
networks~\cite{borgatti00, csermely13,rombach14,zhang15,barucca16, ma15}. However, not all interactions are the
same, and networks whose nodes are connected through connections which
can vary in meaning and nature, can be better described in terms of
networks with many layers~\cite{kivela14, boccaletti14,
  battiston17challenges, dedomenico13, battiston14}.  In this work, we
introduced a method to identify core-periphery structure in multiplex
networks and an algorithmic procedure to extract the multiplex core of
the system. The algorithm was first shown to work on synthetic
multiplex networks with tunable core similarity, i.e. with controlled
overlap between the cores at the different layers. Although the
algorithm is very general and can turn useful in several other
contexts, it was applied here to a specific problem, namely the
identification of the core of the human connectome. 

Finding the router
regions that ensure integration between the different brain modules
and communication in the system as a whole can help answer
fundamental questions in neuroscience. The existence of a network core
in the brain is considered a prerequisite for neural functioning and
cognition, and damages to the core have been recently associated with several
neurological or psychiatric diseases \cite{VDH13,gollo_dwelling_2015,daianu_rich_2015}.  
Previous studies have addressed the question by mainly considering the
structural connectivity of the brain, and by using several techniques,
such as $k$-core decomposition, centrality measures, and rich-club
analysis \cite{hagmann_mapping_2008,heuvel_rich-club_2011}. 
Standard analyses of the structural connectivity of the human brain
agree on the implication of posterior medial and parietal cortical
regions in the network core, but are contradictory on the role of
frontal regions, such as the mPFC, that are functional components of
the default-mode network (DMN)~\cite{buckner_brains_2008}.  
Current trends, however, point out that a better understanding of brain networks can only be obtained by considering 
together the different types of interactions, and this can be achieved
by investigating simultaneously both structural and
functional brain connectivity~\cite{battiston17brain, crofts_structure-function_2016, betzel_multi-scale_2017}.

\begin{figure*}[ht!]
\begin{center}
\includegraphics[width=\textwidth]{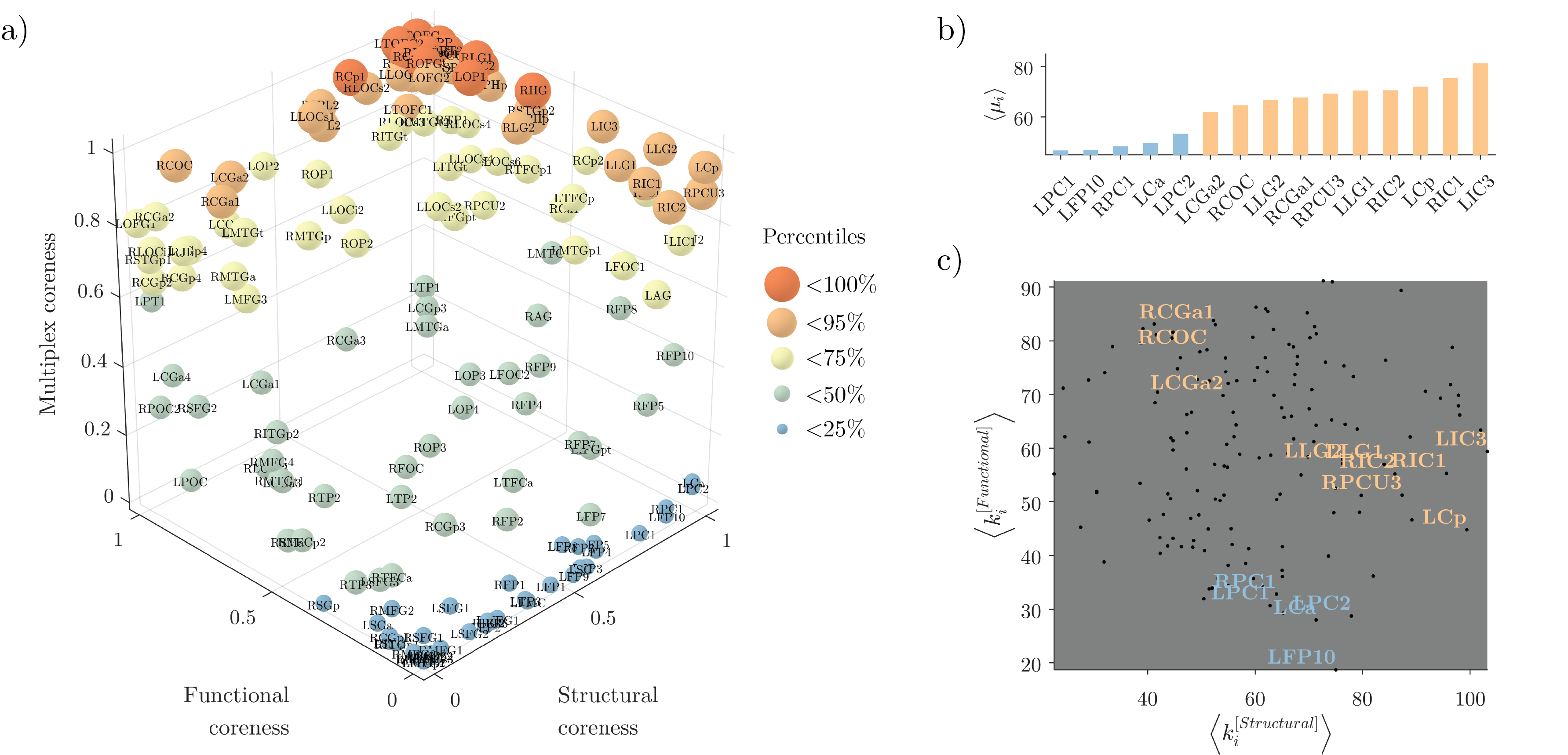}
\caption{\textbf{Emergent non-trivial core regions in the multiplex brain.} 
Panel (a) shows the scatter plot of the structural, functional and multiplex coreness of the regions of interest (ROIs) in the brain. The color and size of the nodes are associated to the percentile of multiplex coreness across the set of brain regions, as in Fig. \ref{fig:fig_kall}.  
    Panel (b) reports the average value of multiplex richness $\langle \mu_i \rangle$ across the different thresholds for the ROIs with the strongest differences in structural and functional coreness.  The color follows the same legend as in panel (a).
    Panel (c) illustrates the distribution of the ROIs (black points) as a  function of their averaged structural and functional degree across all the thresholds. Only the ROIs listed in panel (b) are highlighted according to the same color legend as in panel (a).} 
\label{fig:fig_scatter}
\end{center}
\end{figure*}

Our method allows to elucidate this aspect,
by exploiting also the information available in the
functional connectivity, which has surprisingly been poorly
explored for such a purpose. 
The results we have obtained confirm, on the one hand, the systematic
involvement of posterior medial (C, IC, PCU) and parietal cortices
(e.g., SPL) into the rich core of the human brain. Indeed, these
regions have been already identified through the analysis of the
structural connectome.  On the other hand, the mPFC (e.g., PC and FP),
which exhibited a high structural but low functional coreness, is
eventually assigned to the periphery (Fig. \ref{fig:fig_scatter}a,
lower-right corner).  Notably, this result can be predicted by the
lower multiplex richness and relatively low structural degree, and not
by the solely attitude of frontal areas to be peripheral in the
functional brain network (Fig.~\ref{fig:fig_scatter}b,c).
The exclusion of the mPFC from the rich core supports the hypothesis
that default-mode network activity may be mainly driven from highly
coupled areas of the posterior medial and parietal cortex, which in
turn link to other highly connected regions, such as the medial
orbitofrontal cortex~\cite{hagmann_mapping_2008}.

While frontal ROIs are excluded, new regions gain importance and
become part of the core because of their higher multiplex richness
(see Fig. \ref{fig:fig_scatter}a, upper left corner). Among them, we
report areas of the central gyrus (CGa, CGp to a minor extent), which
were characterized by a low structural but relatively high functional
degree, as shown in Fig. \ref{fig:fig_scatter}(b,c).  These regions
are part of the primary sensori-motor cortex, which has been shown to
be the most extensive of the resting-state components, or networks (out of 8
\cite{heuvel_exploring_2010}), covering $27\%$ percent of
the total gray matter in the brain~\cite{tomasi_association_2011}.
The primary sensori-motor component has a high degree of integration (overlap and activity coupling) with
all other resting-state networks (e.g., DMN), which is consistent with
the increased synchronization of neural activity in cortical regions
during sensory processing \cite{srinivasan_increased_1999} and suggests
an important role in conscious perception.
Notably, ongoing functional connectivity in the primary sensori-motor
network, originally revealed by seed-based analysis \cite{biswal_functional_1995,xiong_interregional_1999}, has been
extensively verified by ICA and clustering methods \cite{salvador_neurophysiological_2005,damoiseaux_consistent_2006}.

The approach we have proposed here provides an effective tool to integrate
mesoscale topological information in brain networks derived from
multimodal neuroimaging data.  Indeed, multimodal integration in
neuroscience is gaining more and more interest due, on the one hand, to
the increasing availability of large heterogenous datasets (e.g. HCP http://www.humanconnectomeproject.org, ADNI http://adni.loni.usc.edu) and, on the other hand, to the need of principled ways to
define more robust biomarkers. Based on a local fast algorithm, our 
method allows to extract the core-periphery structure of multiplex
brain networks, which can be used to characterize multiscale neural
mechanisms (e.g., cross-frequency coupling) and as predictive
diagnostics for multifactor brain diseases, such as Alzheimer's disease.

It is important to notice, that our analysis of the human connectome
relies on the assumption that each layer contributes with the same
intensity to the definition of the multiplex core. In general,
however, the contribution of a layer $\alpha$ can be weighted
differently through an opportune choice of the parameter
$c\lay{\alpha}$, and this can be used to enhance or reduce the
importance of the different types of connectivity.  A larger value of
$c\lay{\alpha}$ increases the relevance of the corresponding layer
until when, in the limit in which $c\lay{\alpha} \to 1$ and the
coefficients of all the other layers go to zero, the multiplex core is
not any more defined by the topology of all the $M$ layers, but
coincides with the core at layer $\alpha$. For instance, setting
$c\lay{structural} = 1$ and $c\lay{functional} = 0$ returns a core
based on the anatomical information only, 
and in agreement with most of the
previous literature on such topic (see
Fig.~\ref{fig:fig_kall_struct}).  As a first attempt to characterize
the multiplex core of the human brain, we decided to
focus our analysis on the simplest and symmetric case, 
$c\lay{structural} = c\lay{functional} = 1/2$, albeit other
combinations are in general possible and can be explored if supported
by a plausible rationale. For example, in the case of multifrequency
brain networks, one could assign stronger weights to higher frequency
layers in order to compensate for $1/f$ frequency scaling of power spectra~\cite{bedard_does_2006}.

To conclude, our method to investigate multiplex core-periphery
organization in networks shows that the core of the human cortex is
made up of known hubs, such as posterior medial and parietal cortical
regions, as well as of hubs that were previously overlooked by
standard single-layer approaches. Examples are sensori-motor
areas. Our findings offer an alternative definition of the rich core
of a network, which takes into account not only the anatomical
structure but also brain function. We hope our work will trigger
additional studies to explore the composition of the multiplex core
using functional connectivity acquisition in task-based experiments,
in an effort to better integrate the one-to-many relationships that
exist between structure and function in the human brain \cite{friston_functional_2011}.

\section{\textbf{Methods}}

\subsection{Multiplex stochastic block model with tunable core similarity}

Modelizations of multiplex networks in terms of stochastic block
models can be found in Ref.~\cite{peixoto15}. Here, we introduce a
stochastic block model that allows to sample multiplex networks with 
an assigned value of core similarity $S_C$ (see Eq.~\ref{eq:S_c}).
Suppose we have $N$ nodes and we want to construct a multiplex network
having a core-periphery structure at each layer $\alpha=1, \ldots, M$,
with $N_c\lay{\alpha}$ nodes in the core of layer $\alpha$.  In
particular, we set $M=2$, $N=250$,  
$N_c\lay{1}=N_c\lay{2}=N_c=50$, and we create at each layer a 
core-periphery structure with the same set of densities:
$\rho_1=0.2$, $\rho_2=0.04$ and $\rho_3=0.03$. Namely, for each of the two layers, 
we connect with a probability $\rho_1$ two nodes both in the core, with 
probability $\rho_2$ a node in the core and a node in the 
periphery, and finally with probability $\rho_3$ two
peripheral nodes.
The values of the
three parameters were chosen in a way that $\langle k \rangle=10$ on
both layers, and the core-periphery structure of each layer is
sufficiently strong to be detected with good accuracy, as discussed in
the Supplementary Information.  Different levels of core similarity are
achieved by varying the overlap between core nodes at the two
layers. When the two sets of core nodes are completely overlapping,
$S_c=1$, whereas when the two sets are disjoint $S_c=0$. Despite other
related formulations of $S_c$ are possible, our definition reflects
the intuition that when two layers with equal core size share half of
the core nodes, then $S_c=1/2$.

\subsection{Multiplex richness $\mu_i$ and $\mu_i^+$}

The multiplex richness $\mu_i$ and $\mu_i^+$ introduced in Eqs.~\ref{eq:mu} and \ref{eq:muplus} are obtained by mean of a simple aggregation of information based on the single layers. In the simplest set-up $c\lay \alpha=c=1/M$ for $\alpha=1, \ldots, M$, and the multiplex richness $\mu_i$ of a node $i$ is simply proportional to its overlapping degree $o_i$~\cite{battiston14}.  A layer with higher density weighs more in the computation of the multiplex core of a network. In general, coefficients $c\lay{\alpha}$ can be used to modulate the relevance to the layers of the network in order to extract its core. If one wants to have equal contributions to $\mu_i$ and $\mu_i^+$ from all the layers but their number of links $K\lay{\alpha}$ is different - for instance because in some layers it might be easier to establish or measure a connection than in others - a natural choice is to set $c\lay{\alpha}$ to be proportional to $1/K\lay{\alpha}$. In other cases, independently from their density, it might be reasonable to assign different importance to different layers, because of exogenous information. Once again this can be achieved by assigning different values of the coefficients $c\lay{\alpha}$ 
At last, we notice that Eq.~\ref{eq:mu} is a particular choice of a more general scenario, where the multiplex richness $\mu_i$ is a generic function $f$ of the degree of a node at the different layers:
\begin{equation}
\mu_i = f(k_i\lay 1, \ldots, k_i\lay M).
\end{equation}
and $\mu_i^+$ is a function of a generic function $g$:
\begin{equation}
\mu_i^+ =g (k_i^{+[1]}, \ldots, k_i^{+[M]}).
\end{equation}

\subsection{Multimodal brain networks}
We considered $171$ healthy human subjects from the NKI Rockland dataset \url{http://fcon_1000.projects.nitrc.org/indi/pro/nki.html}. We used diffusion weighted magnetic resonance imaging (dwMRI) and functional magnetic resonance imaging (fMRI) to derive respectively structural and functional brain networks in each subject.

We gathered the corresponding connectivity matrices from the USC Multimodal Connectivity Database (\url{http://umcd.humanconnectomeproject.org}) \citep{brown_connected_2016}. 

In particular, structural connectivity was obtained using anatomical fiber assignment through the continuous tracking (FACT) algorithm \citep{mori_fiber_2002}.
Functional connectivity was computed by means of the Pearson's correlation coefficient between fMRI signals. More details about the processing steps can be found here \citep{brown_ucla_2012}.
A total number of $N=188$ regions of interest (ROIs)   were available for both structural and functional brain networks, thus resulting in connectivity matrices of size $N \times N$, spatially matched with the MNI152 template \citep{craddock_whole_2012}.

Because we were interested in cortical networks, we removed all subcortical ROIs and obtained connectivity matrices of size $158 \times 158$. The full name and acronym for all the ROIs can be found in Table S1. We then averaged the resulting connectivity matrices (after Fisher transformation) across subjects in order to have a population-level representation.
At the end, we obtained a structural weighted connectivity matrix $\mathcal S$, whose entry $s_{ij}=s_{ji}$ contained the group-average number of axonal fibers between ROIs $i$ and $j$, and a functional weighted connectivity matrix $\mathcal F$, whose entry $f_{ij}=f_{ji}$ corresponded to the group-average correlation coefficient between the fMRI signals of ROIs $i$ and $j$.

We used density-based thresholding to derive structural and functional brain networks by removing the lowest values from the connectivity matrices and binarizing the remaining ones \cite{de_vico_fallani_graph_2014}. We considered a full range of density thresholds, corresponding to an increasing average node degree $\langle k \rangle=1,2,.., 120$. The last value was given by the maximal $\langle k \rangle$ observed in the native structural connectivity matrices, which were originally not fully connected. 
After filtering, for each threshold we combined the resulting structural and functional brain networks into a multiplex network $\mathcal M=\{\mathcal S, \mathcal F \}$.



\clearpage
\beginsupplement
%
%
%
%
%
%


\section*{\textbf{Supplementary information}}

\subsection*{Stochastic block model for rich cores in single-layer networks}

Suppose we have $N$ nodes and we want to construct a single-layer network from which we can identify a partition into two sets: a core of size $N_c<N$ and a periphery of size $Np=N-N_c$. Here we tested the performance of the single-layer algorithm to detect rich cores~\cite{ma15} on a simple stochastic block model.
Let us consider $N$ nodes from which $N_c$ drawn at random are chosen to be part of the network core, whereas the remaining $N_p$ are part of the periphery. A network with core-periphery structure is such that its adjacency matrix can be decomposed into four different blocks: a dense diagonal block  
encoding information on core-core links, a sparser diagonal block describing links among peripheral nodes, 
and two off-diagonal blocks encoding core-periphery edges. In our block model, we connect two nodes with probability $\rho_1$ if they both belong to the core, with probability $\rho_2$ if one of them belongs to the core and one to the periphery, and with probability $\rho_3$ if they both belong to the periphery, $\rho_1 \ge \rho_2 \ge\rho_3$. Given a stochastic realisation of the block model, we can extract the rich core of the network and compare it with the groundtruth, i.e. the set of nodes originally labeled as core nodes. In particular, we can test the accuracy of the algorithm for different choice of the parameters $\rho_1$, $\rho_2$ and $\rho_3$.

Given the three probabilities, the expected total number of edges connecting two core nodes is $K_{cc} = \rho_1 [(N_c-1)*N_c/2]$, the expected total number of edges connecting two peripheral nodes is $K_{pp} = \rho_3 [(N-N_c-1)*(N-N_c)/2]$, and the expected total number of edges connecting a node in the core and a node in the periphery $K_{cp}=\rho_2 [N_c*(N - N_c)]$. The total number of links is $K=K_{cc}+K_{cp}+K_{pp}$.

\begin{figure*}[ht!]
\begin{center}
\includegraphics[width=3in]{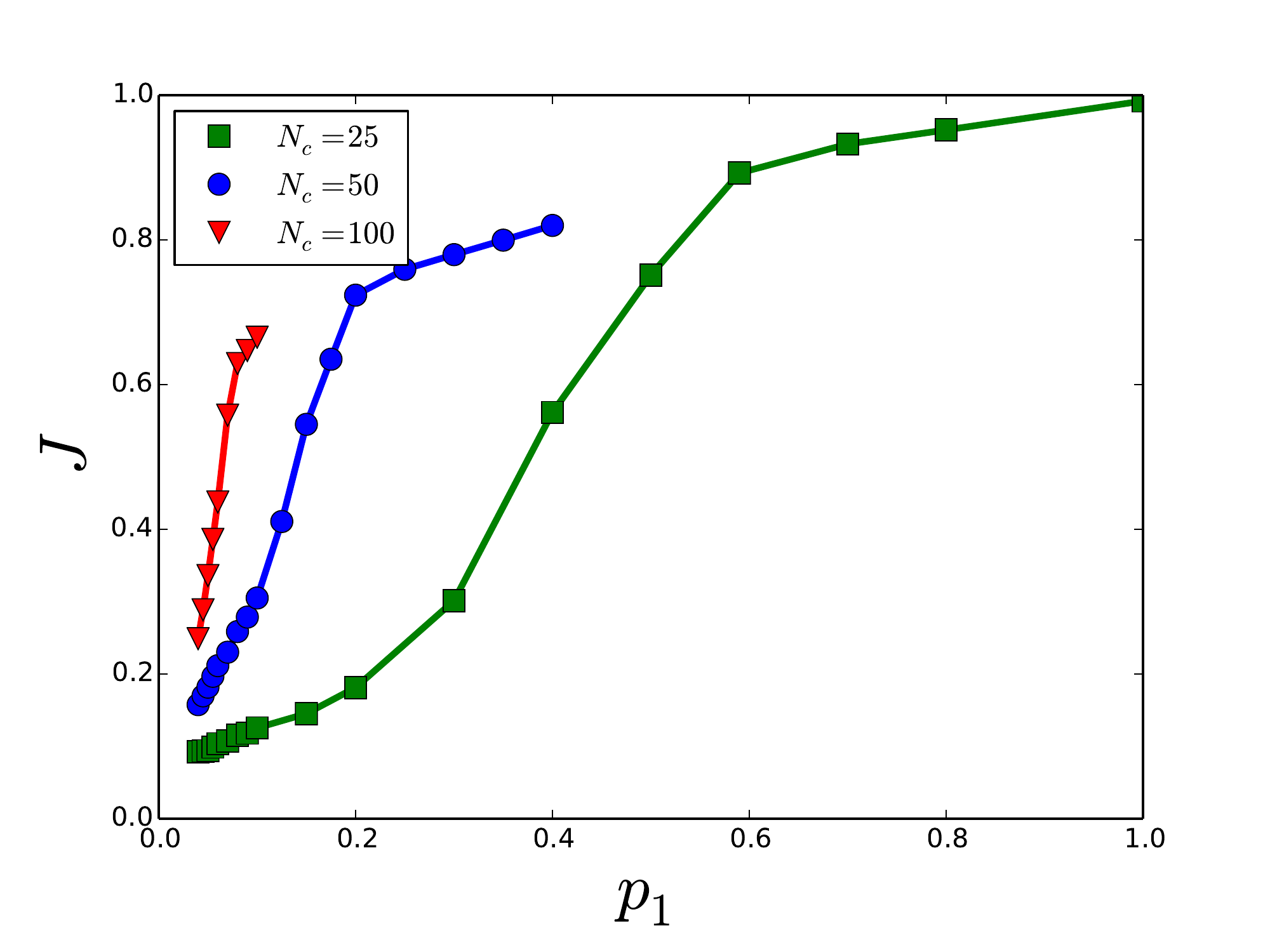}
\caption{Jaccard index $J$ for the groundtruth core-periphery partition and the partition obtained by the algorithm on realisations of the stochastic block model as a function of $\rho_1$ and for different core sizes $N_c$.}
\label{fig:SI_singlelayer_a}
\end{center}
\end{figure*}

In the case $\rho_1 = \rho_2 = \rho_3 = \rho$ the nodes are statistically indistinguishable from a structural point of view, the network lacks a core-periphery structure and specifying the value of $\rho$ simply sets the expected average degree of the network $\langle k \rangle = N \rho$. For instance, for $N=250$ and $\rho=0.04$ we obtain 
$\langle k \rangle=10$ and $K=1250$. Of the different blocks of the adjacency matrix, the exact value of the density of the block encoding links between core and periphery nodes does not play a significant role~\cite{borgatti00}. For such a reason here we set $\rho_2=0.04$, and study the core-periphery structure of the network as a function of $\rho_1$, with $\rho_1>\rho_2$. The higher the value of $\rho_1$, the stronger the core-periphery structure of the system. In order to control for the density of the network, as we increases $\rho_1$ we have to opportunely decrease the value of $\rho_3$. The average degree $\langle k \rangle$ can be kept fixed by setting
\begin{equation}
\rho_3 = \frac{2}{(Np)* (Np - 1)}\biggl(K - K_{cc} - K_{cp} \biggr).
~\label{eq:density3}
\end{equation}
In our case with $N=250$ and $\langle k \rangle=10$, we have $K=1250$ whereas $K_{cc}$ and $K_{cp}$ are set once we fix the core size $N_c$ and the value of $\rho_1$. In Fig.~\ref{fig:SI_singlelayer_a} we show the average Jaccard index $J$ computed for the groundtruth partition and the partition extracted by the algorithm on the stochastic realisations of the network as a function of different values of $\rho_1$ for different core size. As shown, $J$ increases quickly until $\rho_1=0.2$ and only mildly after this point. This indicates that $\rho_1=0.2$, corresponding to a value of $\rho_3=0.03$, can be considered as the smallest density of the core-core block at which the core-periphery structure of the network is sufficiently well-defined. For this reason, in the stochastic block model for multiplex networks with different values of core similarity $S_c$ described in Fig.~\ref{fig:fig_sim}, where we have $N=250$ and $N_c=50$ we set $\rho_1=0.2$.  

Given the set of parameters $\rho_1$, $\rho_2$ and $\rho_3$ we can also compute the average degree $\langle k_c \rangle$ of core nodes
\begin{equation}
\langle k_c \rangle = \rho_1 (N_c -1) + \rho_2 (N_p),
\end{equation}
the average degree $\langle k_p \rangle$ of the peripheral nodes
\begin{equation}
\langle k_p \rangle = \rho_3 (N_p -1) + \rho_2 (N_c).
\end{equation}
so that we have
\begin{equation}
\langle k \rangle = \frac{N_c\langle k_c \rangle + N_p \langle k_p \rangle}{N}.
\end{equation}

\begin{figure*}[ht!]
\begin{center}
\includegraphics[width=3in]{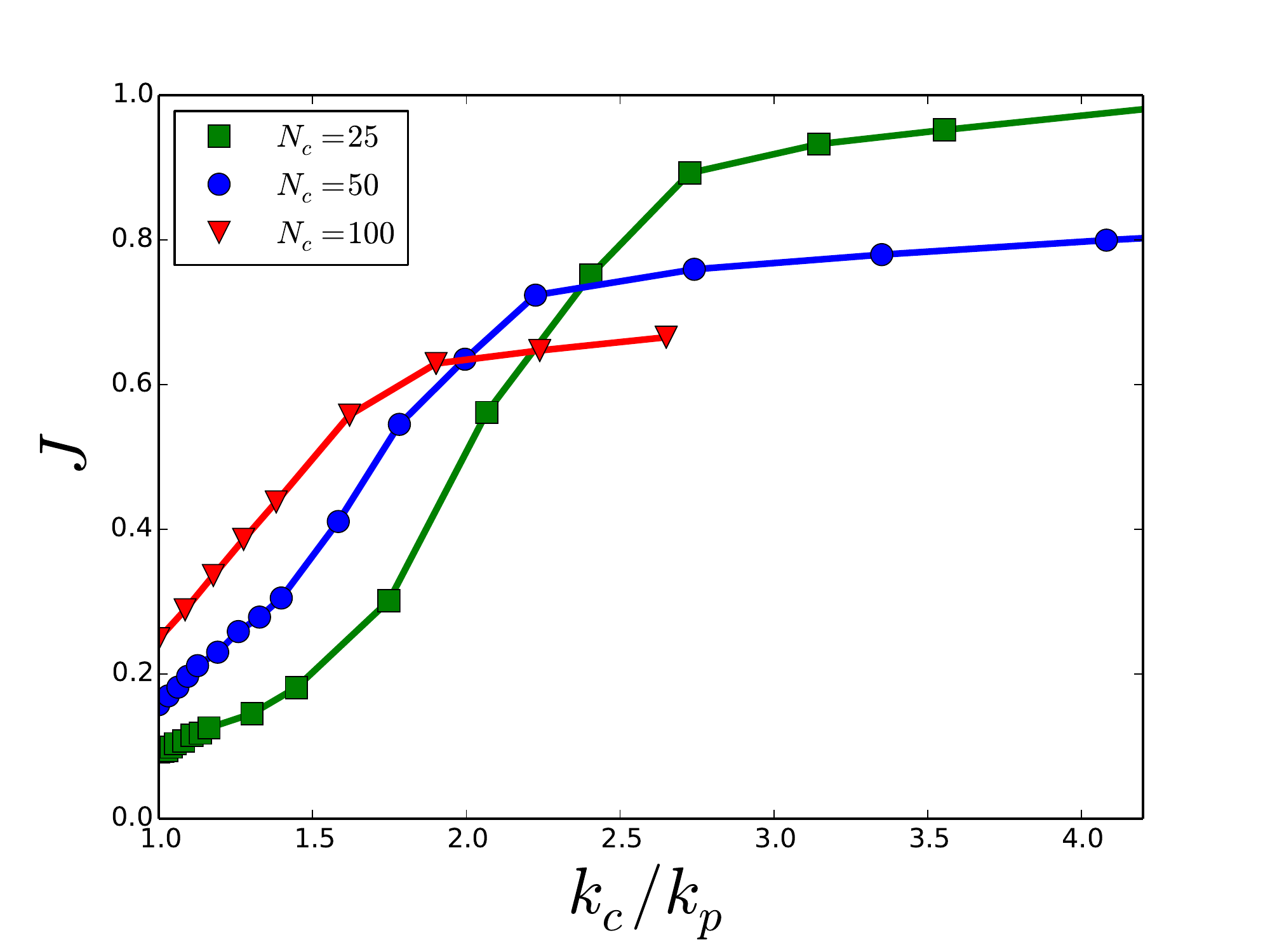}
\caption{Jaccard index $J$ for the groundtruth core-periphery partition and the partition obtained by the algorithm on realisations of the stochastic block model as a function of $\langle k_c \rangle/\langle k_p \rangle$ and for different core sizes $N_c$.}
\label{fig:SI_singlelayer_b}
\end{center}
\end{figure*}

In Fig.~\ref{fig:SI_singlelayer_b} we show the average Jaccard index $J$ computed for the groundtruth partition and the partition extracted by the algorithm as a function of $\langle k_c \rangle/\langle k_p \rangle$.

\clearpage

\subsection*{Cores of the Top Noordin Terrorists network}

In the first Table we report the size $N_c$ of the cores of the three layers (mutual trust, common operations, exchanged communications) of the Top Noordin Terrorists network~\cite{battiston14} and of the multiplex core shown in Fig.~\ref{fig:fig_terr}. 
\begin{center}
 \begin{tabular}{||c | c||} 
 \hline
 \textbf{Layer} & \textbf{$N_c$} \\ 
 \hline\hline\hline
  1 & 17 \\ 
 \hline
  2 & 17 \\
 \hline
  3 & 12 \\
 \hline
 \hline
  Multiplex & 12  \\ 
  \hline
\end{tabular}
\label{table:noordin1}
\end{center}

In the second Table we report the number of common core nodes $I_c$ belonging to the different pairs of layers. The network is characterised by a core similarity  $S_c=0.38$ ($S_c\lay1=0.32$, $S_c\lay2=0.35$, $S_c\lay1=0.46$. See Eq.~\ref{eq:S_c} in the main text). We also report the number of common core nodes for the multiplex and each layer. 
\begin{center}
 \begin{tabular}{||c | c | c||} 
 \hline
 \textbf{Layer} & \textbf{Layer} & \textbf{$I_C$} \\ 
 \hline\hline\hline
  1 & 2 &  6\\ 
 \hline
  1 & 3 & 5 \\
 \hline
  2 & 3 & 6 \\
 \hline
 \hline
  Multiplex & 1 & 10  \\ 
  \hline
  Multiplex & 2 & 8  \\ 
  \hline
  Multiplex & 3 & 7  \\ 
  \hline
\end{tabular}
\label{table:noordin2}
\end{center}

\subsection*{Core similarity for structural and functional brain networks}
\begin{figure*}[ht!]
\begin{center}
\includegraphics[width=3in]{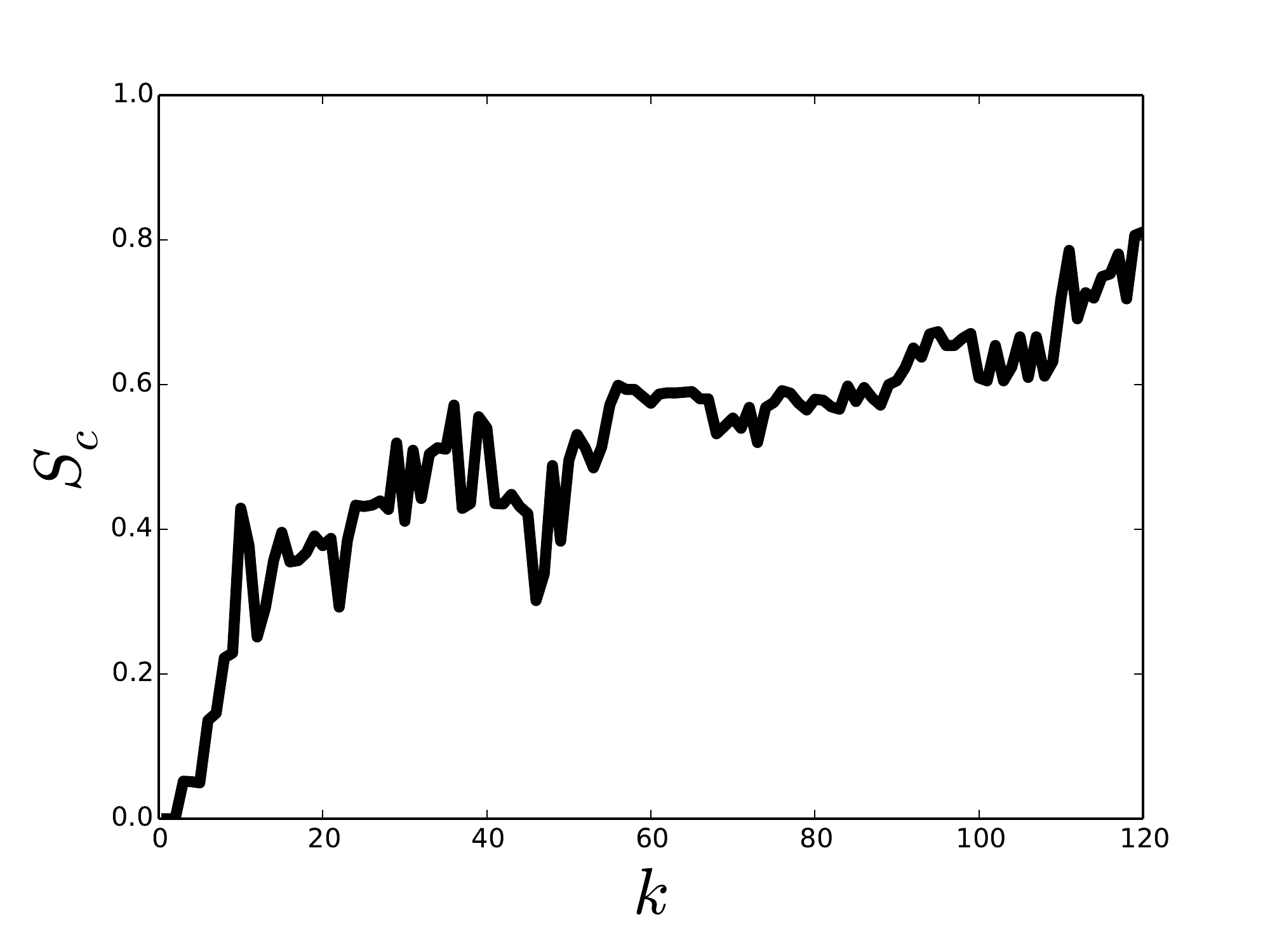}
\caption{Core similarity $S_c$ for the structural and functional networks thresholded at different values of average degree $\langle k \rangle$.}
\label{fig:SI_braincoresimilarity}
\end{center}
\end{figure*}

In Fig.~\ref{fig:SI_braincoresimilarity} we show the core similarity $S_c$ for the considered averaged structural and functional networks as a function of different thresholds.

\clearpage

\subsection*{Structural and functional coreness in the human brain}

\begin{figure*}[ht!]
\begin{center}
\includegraphics[width=7in]{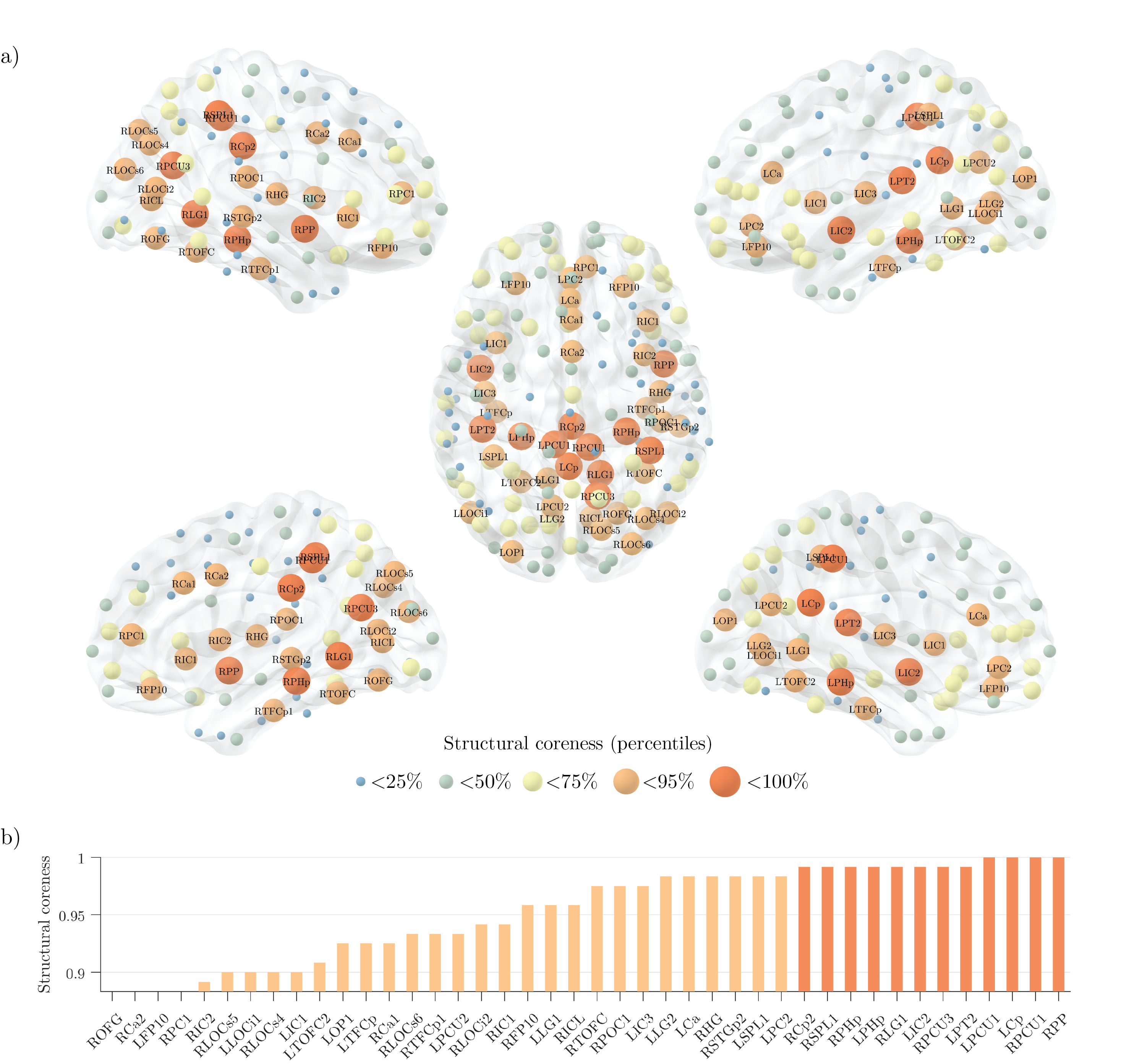}
\caption{\textbf{Structural core-periphery structure of the human cortex.} In panel (a) we show the node structural coreness from different points of view: external view in the top row, internal view in the bottom row. The color and size of each node code for the percentile to which it belongs as specified in the legend. In panel (b) we report the value of structural coreness of the nodes beyond the 75th percentile with the same color code.}
\label{fig:fig_kall_struct}
\end{center}
\end{figure*}

\begin{figure*}[ht!]
\begin{center}
\includegraphics[width=7in]{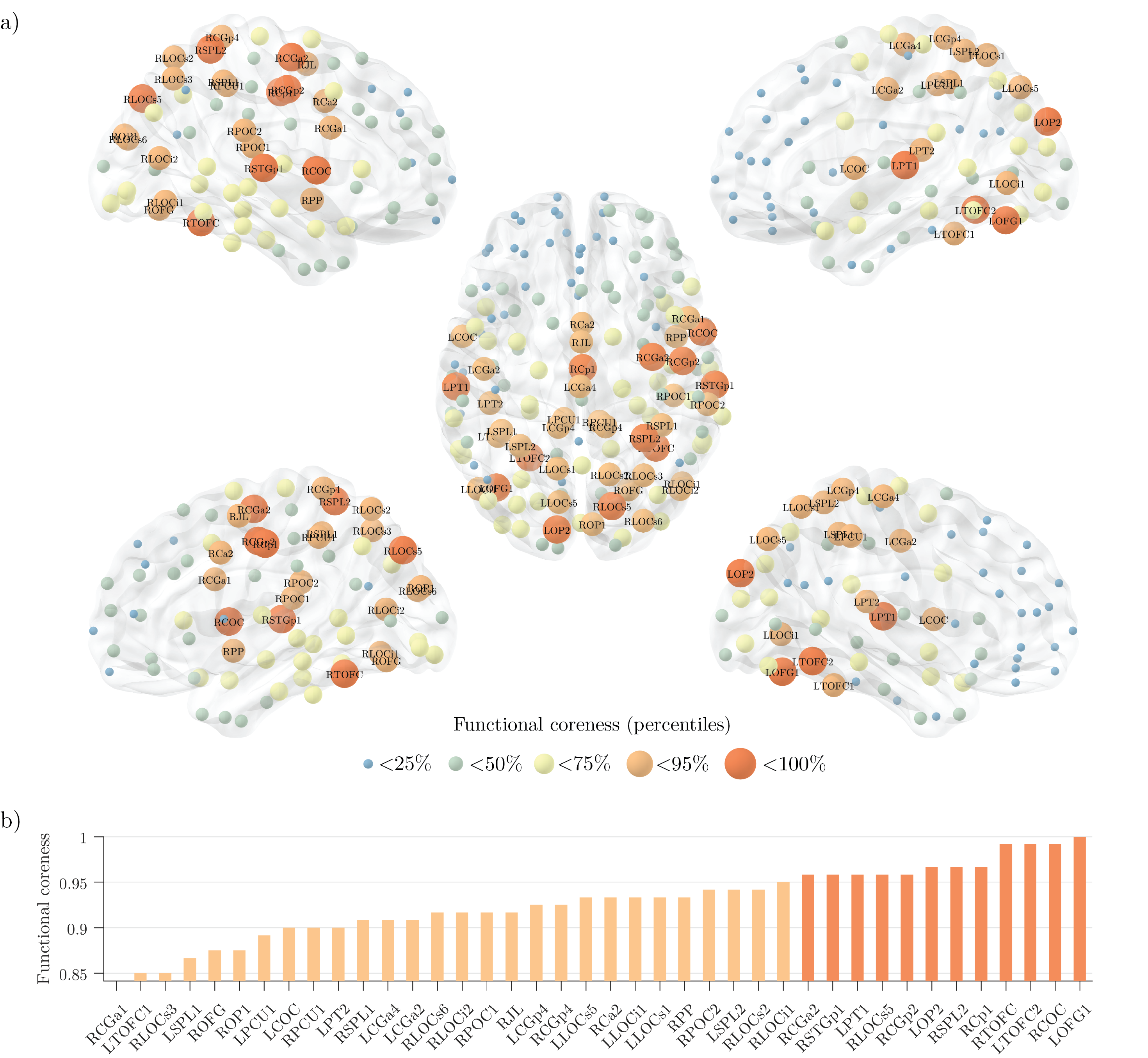}
\caption{\textbf{Functional core-periphery structure of the human cortex.} In panel (a) is represented the functional coreness from different points of view: external view in the top row, internal view in the bottom row. The color and size of each node code for the percentile to which it belongs as specified in the legend. In panel (b) we report the value of functional coreness for the nodes beyond the 75th percentile with the same color code.}
\label{fig:fig_kall_func}
\end{center}
\end{figure*}

In Figs.~\ref{fig:fig_kall_struct} and ~\ref{fig:fig_kall_func} we report for the ROIs of the human brain the node coreness computed respectively at the structural and functional layer.  

\subsection*{List of ROIs}

The full list of the considered Regions of Interested (ROIs), and the corresponding abbreviations, can be found in the following Table S1.

\clearpage 

\begin{table}[h!]
\scriptsize
  \begin{center}
    
    \label{tab:table1}
    \begin{tabular}{|l|l||l|l|} 
    \hline
      \textbf{Label} & \textbf{Abbreviation} & \textbf{Label} & \textbf{Abbreviation}\\
      \hline
      \hline
Left Angular  &  LAG  &     Right Central Opercular  &     	RCOC\\
Left Central Opercular  &     	LCOC  &     	Right Cingulate anterior 1	  &     RCa1\\
Left Cingulate anterior  &     	LCa  &     	Right Cingulate anterior 2  &     	RCa2\\
Left Cingulate posterior	  &     LCp  &     	Right Cingulate posterior 1  &     	RCp1\\
Left Frontal Medial  &     	LFMC  &     	Right Cingulate posterior 2	  &     RCp2\\
Left Frontal Orbital 1  &     	LFOC1  &     	Right Frontal Orbital  &     	RFOC\\
Left Frontal Orbital 2	  &     LFOC2  &     	Right Frontal Pole 1  &     	RFP1\\
Left Frontal Pole 1  &     	LFP1  &     	Right Frontal Pole 10  &     	RFP10\\
Left Frontal Pole 10  &     	LFP10  &     	Right Frontal Pole 2  &     	RFP2\\
Left Frontal Pole 2  &     	LFP2  &     	Right Frontal Pole 3  &     	RFP3\\
Left Frontal Pole 3  &     	LFP3  &     	Right Frontal Pole 4  &     	RFP4\\
Left Frontal Pole 4  &     	LFP4  &     	Right Frontal Pole 5  &     	RFP5\\
Left Frontal Pole 5  &     	LFP5  &     	Right Frontal Pole 6  &     	RFP6\\
Left Frontal Pole 6  &     	LFP6  &     	Right Frontal Pole 7  &     	RFP7\\
Left Frontal Pole 7  &     	LFP7  &     	Right Frontal Pole 8  &     	RFP8\\
Left Frontal Pole 8  &     	LFP8  &     	Right Frontal Pole 9  &     	RFP9\\
Left Frontal Pole 9  &     	LFP9  &     	Right Heschls  &     	RHG\\
Left Inferior Frontal pars triangularis  &     	LIFGpt  &     	Right Inferior Frontal pars triangularis  &     	RIFGpt\\
Left Inferior Temporal posterior 1	  &     LITGp1  &     	Right Inferior Temporal posterior 1  &     	RITGp1\\
Left Inferior Temporal posterior 2	  &     LITGp2  &     	Right Inferior Temporal posterior 2  &     	RITGp2\\
Left Inferior Temporal temporooccipital  &     	LITGt  &     	Right Inferior Temporal temporooccipital  &     	RITGt\\
Left Insular 1  &     	LIC1  &     	Right Insular 1  &     	RIC1\\
Left Insular 2  &     	LIC2  &     	Right Insular 2  &     	RIC2\\
Left Insular 3  &     	LIC3  &     	Right Intracalcarine  &     	RICL\\
Left Lateral Occipital inferior 1  &     	LLOCi1  &     	Right Juxtapositional Lobule  &     	RJL\\
Left Lateral Occipital inferior 2  &     	LLOCi2  &     	Right Lateral Occipital inferior 1	  &     RLOCi1\\
Left Lateral Occipital superior 1	  &     LLOCs1  &     	Right Lateral Occipital inferior 2	  &     RLOCi2\\
Left Lateral Occipital superior 2  &     	LLOCs2  &     	Right Lateral Occipital inferior 3  &     	RLOCi3\\
Left Lateral Occipital superior 3  &     	LLOCs3  &     	Right Lateral Occipital superior 1  &     	RLOCs1\\
Left Lateral Occipital superior 4	  &     LLOCs4  &     	Right Lateral Occipital superior 2  &     	RLOCs2\\
Left Lateral Occipital superior 5  &     	LLOCs5  &     	Right Lateral Occipital superior 3  &     	RLOCs3\\
Left Lateral Occipital superior 6  &     	LLOCs6  &     	Right Lateral Occipital superior 4  &     	RLOCs4\\
Left Lingual 1  &     	LLG1  &     	Right Lateral Occipital superior 5  &     	RLOCs5\\
Left Lingual 2  &     	LLG2  &     	Right Lateral Occipital superior 6  &     	RLOCs6\\
Left Middle Frontal 1  &     	LMFG1  &     	Right Lingual 1	  &     RLG1\\
Left Middle Frontal 2  &     	LMFG2  &     	Right Lingual 2	  &     RLG2\\
Left Middle Frontal 3	  &     LMFG3  &     	Right Middle Frontal 1  &     	RMFG1\\
Left Middle Temporal anterior  &     	LMTGa  &     	Right Middle Frontal 2  &     	RMFG2\\
Left Middle Temporal posterior 1  &     	LMTGp1  &     	Right Middle Frontal 3  &     	RMFG3\\
Left Middle Temporal posterior 2  &     	LMTGp2  &     	Right Middle Frontal 4  &     	RMFG4\\
Left Middle Temporal temporooccipital  &     	LMTGt  &     	Right Middle Temporal anterior  &     	RMTGa\\
Left Occipital Fusiform 1  &     	LOFG1  &     	Right Middle Temporal posterior  &     	RMTGp\\
Left Occipital Fusiform 2  &     	LOFG2  &     	Right Middle Temporal temporooccipital 1  &     	RMTGt1\\
Left Occipital Pole 1  &     	LOP1  &     	Right Middle Temporal temporooccipital 2	  &     RMTGt2\\
Left Occipital Pole 2	  &    LOP2  &     	Right Occipital Fusiform  &     	ROFG\\
Left Occipital Pole 3  &     	LOP3  &     	Right Occipital Pole 1  &     	ROP1\\
Left Occipital Pole 4  &     	LOP4  &     	Right Occipital Pole 2  &     	ROP2\\
Left Paracingulate 1  &     	LPC1  &     	Right Occipital Pole 3  &     	ROP3\\
Left Paracingulate 2  &     	LPC2  &     	Right Paracingulate 1  &     	RPC1\\
Left Parahippocampal posterior  &     	LPHp  &     	Right Paracingulate 2  &     	RPC2\\
Left Parietal Operculum  &     	LPOC  &     	Right Parahippocampal posterior  &     	RPHp\\
Left Planum Temporale 1  &     	LPT1  &     	Right Parietal Operculum 1  &     	RPOC1\\
Left Planum Temporale 2  &     	LPT2  &     	Right Parietal Operculum 2  &     	RPOC2\\
Left Postcentral 1  &     	LCGp1  &     	Right Planum Polare	  &     RPP\\
Left Postcentral 2  &     	LCGp2  &     	Right Postcentral 1  &     	RCGp1\\
Left Postcentral 3  &     	LCGp3  &     	Right Postcentral 2  &     	RCGp2\\
Left Postcentral 4  &     	LCGp4  &     	Right Postcentral 3  &     	RCGp3\\
Left Precentral 1  &     	LCGa1  &     	Right Postcentral 4  &     	RCGp4\\
Left Precentral 2  &     	LCGa2  &     	Right Precentral 1  &     	RCGa1\\
Left Precentral 3  &     	LCGa3  &     	Right Precentral 2  &     	RCGa2\\
Left Precentral 4  &     	LCGa4  &     	Right Precentral 3  &     	RCGa3\\
Left Precuneous 1  &     	LPCU1  &     	Right Precuneous 1  &     	RPCU1\\
Left Precuneous 2  &     	LPCU2  &     	Right Precuneous 2  &     	RPCU2\\
Left Subcallosal  &     	LSC  &     	Right Precuneous 3  &     	RPCU3\\
Left Superior Frontal 1  &     	LSFG1  &     	Right Superior Frontal 1  &     	RSFG1\\
Left Superior Frontal 2  &     	LSFG2  &     	Right Superior Frontal 2  &     	RSFG2\\
Left Superior Frontal 3  &     	LSFG3  &     	Right Superior Parietal Lobule 1  &     	RSPL1\\
Left Superior Parietal Lobule 1  &     	LSPL1  &     	Right Superior Parietal Lobule 2  &     	RSPL2\\
Left Superior Parietal Lobule 2  &     	LSPL2  &     	Right Superior Temporal posterior 1  &     	RSTGp1\\
Left Supramarginal anterior  &     	LSGa  &     	Right Superior Temporal posterior 2  &     	RSTGp2\\
Left Supramarginal posterior  &     	LSMp  &     	Right Supramarginal anterior  &     	RSMa\\
Left Temporal Fusiform anterior  &     	LTFCa  &     	Right Supramarginal posterior  &     	RSGp\\
Left Temporal Fusiform posterior  &     	LTFCp  &     	Right Temporal Fusiform anterior  &     	RTFCa\\
Left Temporal Occipital Fusiform 1  &     	LTOFC1  &     	Right Temporal Fusiform posterior 1  &     	RTFCp1\\
Left Temporal Occipital Fusiform 2  &     	LTOFC2  &     	Right Temporal Fusiform posterior 2  &     	RTFCp2\\
Left Temporal Pole 1  &     	LTP1	  &     Right Temporal Occipital Fusiform  &     	RTOFC\\
Left Temporal Pole 2  &     	LTP2  &     	Right Temporal Pole 1  &     	RTP1\\
Left Temporal Pole 3	  &     LTP3  &     	Right Temporal Pole 2  &     	RTP2\\
Right Angular  &     	RAG  &     	Right Temporal Pole 3  &     	RTP3\\
\hline
    \end{tabular}   
  \end{center}
  \caption{Full list of  Regions of Interest (ROIs) and abbreviations. Numbers denote the relative position within a macro area, i.e. higher values stand for more posterior ROIs.}
\end{table}


\end {document}